\documentstyle[preprint,aps]{revtex}

\begin{document}
\draft
\author{Yongge Ma}
\address{Department of Physics, Beijing NormalUniversity,\\
Beijing 100875, China.}
\author{Canbin Liang}
\address{Center of Theoretical Physics, CCAST (World Laboratory),\\
Beijing 100080, China\\
and Department of Physics, Beijing Normal University,\\
Beijing 100875, China.}
\author{Zhiquan Kuang}
\address{Center of Theoretical Physics, CCAST (World Laboratory),\\
Beijing 100080, China\\
and Institute of Mathematics, Academia Sinica, Beijing 100080, China}
\title{On the degenerate phase boundaries }
\maketitle

\begin{abstract}
The structure of the phase boundary between degenerate and nondegenerate
regions in Ashtekar's gravity has been recently studied by Bengtsson and
Jacobson who conjectured that the phase boundary should be always null. In
this paper, we reformulate the reparametrization procedure in the mapping
language and distinguish a phase boundary $\partial M_1$ from its image $%
\phi [\partial M_1]$. It is shown that $\phi [\partial M_1]$ has to be null,
while the nullness of $\partial M_1$ requires some more suitable criterion.
\end{abstract}

\pacs{PACS Numbers: 0420C, 0420F, 0460}

\section{Introduction}

It is well known that Ashtekar's formulation of gravity admits degenerate
triads and hence degenerate metrics [1]. Various kinds of degenerate
solutions to the Ashtekar's equations have been investigated [2-8]. Using a
``covariant approach'', Bengtsson and Jacobson [6] obtained a few
4-dimensional spacetimes containing a ``phase boundary'' separating a
degenerate region from a nondegenerate one.

According to Ref.[6], the covariant approach starts from a nondegenerate
metric which solves Einstein's equations, and then reparametrize one of the
coordinates. This reparametrization is chosen so that it is not a
diffeomorphism at some particular value of the coordinate. Adopting the new
coordinate, the solution can be smoothly matched to a solution to the
Ashtekar equations with a degenerate metric at the surface where the
transformation misbehaves. To make things clearer we reformulate this
procedure as follows. Let $M$ be a 4-dimensional manifold and $M_1$ a
4-dimensional submanifold with a 3-dimensional boundary $\partial M_1$.
Suppose $\hat{M}$ is a 4-dimensional manifold with a nondegenerate metric $%
\hat{g}_{\mu \nu }$ which solves the Einstein's equations, and $\phi $ is a
diffeomorphism from $M_1$ to some open set $\hat{M}_1\subset \hat{M}$.
Extend the domain of $\phi $ to the whole of $M$ so that $\phi :M\rightarrow 
\hat{M}$ is smooth with $M-M_1$ being mapped onto $\phi [\partial M_1]$, and
the pushforward $\phi _{*}$ restricted to the tangent bundle of $\partial $$%
M_1$ to that of $\phi [\partial M_1]$ is nondegenerate. (It is assumed that $%
\phi :M_1\rightarrow \hat{M}_1$ has been chosen so that such an extension is
possible.) Then the pullback $g_{\mu \nu }\equiv \phi ^{*}\hat{g}_{\mu \nu }$
is nondegenerate on $M_1$ and degenerate on $M-M_1$. One therefore has a
spacetime $(M,g_{\mu \nu })$ with a ``phase boundary'' separating a
nondegenerate region from a degenerate one. It is clear that the
``reparametrization procedure'' mentioned above is a special case of this
treatment. The authors of Ref.[6] viewed $\phi [\partial M_1]$ as the phase
boundary and raised an interesting question: Is the phase boundary always
null? They conjectured that the answer is ``yes'' provided that the metric
is a ``regular'' solution to Ashtekar's evolution equations, that is,
solutions in which the canonical variables $(A_a^i,E_i^a)$, the shift vector 
$N^i$, and the lapse density $\underline{N}$ all take finite values which,
except for $\underline{N}$, are allowed to vanish. (Since $\underline{N}=N/%
\sqrt{q}$, where $N$ is the usual lapse scalar and $q$ the determinant of
the spatial metric, the requirement that $\underline{N}$ should stay finite
is non-trivial when the spatial metric becomes degenerate.)

Having reformulated the ``reparametrization procedure'' in the mapping
language as stated above, in our opinion it seems reasonable that the
``phase boundary'' should refer to $\partial M_1$ rather than $\phi
[\partial M_1]$ since the latter is not at all a boundary between a
degenerate region and a nondegenerate one, although Ref.[6] took a different
view. We will first show in Sec.2 that $\phi [\partial M_1]$ has to be a
null hypersurface. In Sec.3 we will argue that under certain circumstances $%
\partial M_1$ could be nonnull as judged by the criterion similar to that of
Ref.[6]. Some discussions about the criterion are given in Sec.4.

\section{On the boundary $\phi [\partial M_1]$}

We now show that the hypersurface $\phi [\partial M_1]$, which is viewed as
the degenerate phase boundary in Ref.[6], must be null if the pullback
metric $g_{\mu \nu }$ on $M$ is a regular solution to Ashtekar's equations.

Consider a ``3+1 decomposition'' of the metric : 
\begin{equation}
ds^2=g_{00}dt^2+2g_{0i}dtdx^i+g_{ij}dx^idx^j=(-N^2+N^iN_i)dt^2+2N_idtdx^i+g_{ij}dx^idx^j,
\end{equation}
where $N$ is the lapse scalar and $N^i$ the shift vector which relates to
the metric components via 
\begin{equation}
g_{ij}N^j=g_{0i},\qquad i=1,2,3.
\end{equation}
Since $q\equiv \det (g_{ij})=0$ in the degenerate region of $M$, there
exists a non-vanishing 3-vector $\lambda ^i$ such that $g_{ij}\lambda ^i=0$,
and Eq.(2) then implies that $g_{0i}\lambda ^i=0$. Hence there exists a
4-vector 
\[
T^\nu =\left( 
\begin{array}{c}
0 \\ 
\lambda ^i
\end{array}
\right) 
\]
at each point of $M-M_1$ such that $g_{\mu \nu }T^\nu =0$. Furthermore, in
the degenerate region the lapse scalar $N$ must vanish in order to keep the
lapse density $\underline{N}$ finite, hence it follows from Eq.(1) that 
\begin{equation}
-g_{00}+g_{0i}N^i=N^2=0.
\end{equation}
Eq.(3) together with Eq.(2) provides another 4-vector 
\[
S^\nu =\left( 
\begin{array}{c}
1 \\ 
-N^i
\end{array}
\right) 
\]
at each point of $M-M_1$ such that $g_{\mu \nu }S^\nu =0$. It is obvious
that $T^\nu $ and $S^\nu $ are linearly independent of each other, and hence
represent two independent degenerate directions of $g_{\mu \nu }$. That is
to say, the degenerate subspace of the tangent space at each point of $M-M_1$
is at least 2-dimensional. Since $\partial M_1$ is 3-dimensional, there must
be some degenerate vector field, $W^\nu $, that is tangent to $\partial M_1$%
. It then follows from the nondegeneracy of the pushforward $\phi _{*}$
(restricted to $\partial M_1$) that there is a vector field, $\phi _{*}W^\nu 
$, on $\phi [\partial M_1]$ such that (i) $\phi _{*}W^\nu \neq 0$; (ii) $%
\phi _{*}W^\nu $ is tangent to $\phi [\partial M_1]$; (iii) $\phi _{*}W^\nu $
is orthogonal to all vector fields tangent to $\phi [\partial M_1]$ due to $%
g_{\mu \nu }=\phi ^{*}\hat{g}_{\mu \nu }$. We therefore conclude that $\phi
[\partial M_1]$ is a null hypersurface with null normal $\phi _{*}W^\nu $,
and hence the conjecture in Ref.[6] has been proved.

Note, however, that the Ashtekar's equations are not at all needed in our
proof. It turns out that these equations being necessary for the validity of
the conjecture as claimed in Ref.[6] is simply caused by an error, i.e., a
superfluous term, $H_R\dot{R}^2$, in Eq.(55) of it.

\section{On the degenerate phase boundary $\partial M_1$}

In this section we will argue through an example that, although $\phi
[\partial M_1]$ is always null, it is not the case for $\partial M_1$
according to the criterion similar to that of Ref.[6]. Since the metric $%
g_{\mu \nu }$ is degenerate on $\partial M_1$, it is a delicate issue what
definition of nullness of $\partial M_1$ is used. Noticing the criterion for
the nullness of $\phi [\partial M_1]$ used in Ref.[6], we define $\partial
M_1$ to be null if it is null ``when viewed from the nondegenerate side''.
More precisely, suppose $\partial M_1$ is given by $f=0$, where $f$ is a
smooth function with $\nabla _af|_{\partial M_1}\neq 0$, then $\partial M_1$
is said to be null if $g^{\mu \nu }\nabla _\mu f\nabla _\nu f\rightarrow 0$
as $\partial M_1$ is approached.

Let $(U,X^i)\,(i=1,2,3)$ be a coordinate system on $\hat{M}$ with $U=0$
representing the null hypersurface $\phi [\partial M_1]$ (assuming that $%
\phi [\partial M_1]$ can be covered by a single 4-dimensional coordinate
patch), and the line element of $\hat{g}_{\mu \nu }$ in this coordinate
system reads 
\begin{equation}
d\hat{s}=\hat{g}_{UU}dU^2+2\hat{g}_{Ui}dUdX^i+\hat{q}_{ij}dX^idX^j.
\end{equation}
The nullness of $\phi [\partial M_1]$ then implies $\lim {_{U\rightarrow 0}}%
\hat{q}=0$ where $\hat{q}\equiv \det (\hat{q}_{ij})$. The mapping $\phi
:M\rightarrow \hat{M}$ induces four functions $\phi ^{*}U,\phi
^{*}X^i\,(i=1,2,3)$ on $M$ with $\phi ^{*}U|_{M-M_1}=0$. Let $(u,x^i)$ be a
coordinate system on $M$ with $u|_{\partial M_1}=0$ and $x^i=\phi ^{*}X^i$,
then one has a function $U(u)$ [short for $(\phi ^{*}U)(u)$ ] with $%
U^{\prime }(u)|_{M-M_1}\equiv \frac{dU}{du}|_{M-M_1}=0$. The line element of 
$g_{\mu \nu }$ in this coordinate system is as follows: 
\begin{equation}
ds^2=U^{\prime }\hat{g}_{UU}du^2+2U^{\prime }\hat{g}_{Ui}dudx^i+\hat{q}%
_{ij}dx^idx^j.
\end{equation}
This is exactly the procedure of ``reparametrization of one of the
coordinates'' mentioned in Sec.1. Now the key quantity needed for judging
whether $\partial M_1$ is null is 
\begin{equation}
g^{\mu \nu }\nabla _\mu u\nabla _\nu u=\hat{q}/g,
\end{equation}
where $g$ denotes the determinant of the line element (5) and can be
expressed as 
\begin{equation}
g=(U^{\prime })^2\hat{g}
\end{equation}
with $\hat{g}$ the determinant of the line element (4), which does not
vanish since $\hat{g}_{\mu \nu }$ is nondegenerate. It then follows from
Eqs.(6) and (7) that 
\begin{equation}
\lim_{u\rightarrow 0^{+}}g^{\mu \nu }\nabla _\mu u\nabla _\nu
u=\lim_{u\rightarrow 0^{+}}\frac{\hat{q}}{(U^{\prime })^2\hat{g}}%
=\lim_{u\rightarrow 0^{+}}\frac 1{2\hat{g}U^{\prime \prime }}\frac{\partial 
\hat{q}}{\partial U},
\end{equation}
where $U^{\prime \prime }\equiv dU^{\prime }/du$, and we assume $u>0$ in $%
M_1 $ for convenience. Since $\hat{g}$ is finite and $U^{\prime \prime }$
approaches zero as $u\rightarrow 0^{+}$, it is quite probable to construct
an example in which the hypersurface $u=0$ is nonnull by requiring $%
\lim_{u\rightarrow 0^{+}}\partial \hat{q}/\partial U\neq 0$ or the rate of
approaching zero of $\partial \hat{q}/\partial U$ is equal to or less than
that of $U^{\prime \prime }$. The following is a concrete example.

Let $(\hat{M},\hat{g}_{\mu \nu })$ be the Minkowski spacetime and the line
element in double null coordinates $(\bar{U},\bar{V},Y,Z)$ reads 
\begin{equation}
d\hat{s}^2=-d\bar{U}d\bar{V}+dY^2+dZ^2.
\end{equation}
A simple coordinate transformation 
\[
\bar{U}=Ue^{-V},\ \bar{V}=V 
\]
turns it to 
\begin{equation}
d\hat{s}^2=-e^{-V}(dU-UdV)dV+dY^2+dZ^2.
\end{equation}
It is obvious that $U=0$ is a null hypersurface which serves as $\phi
[\partial M_1]$ of the previous discussion. Define $u$ on $M_1$ such that 
\begin{equation}
U(u)=u^3e^u
\end{equation}
in $M_1$. [The fact $U(u)=0$ in $M-M_1$ follows automatically from the
mapping $\phi $ that requires $\phi (M-M_1)=\phi [\partial M_1]$.] the
metric $g_{\mu \nu }\equiv \phi ^{*}\hat{g}_{\mu \nu }$ now reads 
\begin{equation}
ds^2=-e^{-V}(U^{\prime }du-UdV)dV+dY^2+dZ^2,
\end{equation}
where 
\begin{equation}
U^{\prime }\equiv \frac{dU}{du}=\left\{ 
\begin{array}{ll}
u^2(u+3)e^u & \mbox{ in }M_1 \\ 
0 & \mbox{ in }M-M_1
\end{array}
\right. .
\end{equation}
It is obvious from Eq.(10) that $\hat{q}=Ue^{-V}$ and hence 
\[
\lim_{u\rightarrow 0^{+}}\frac{\partial \hat{q}}{\partial U}=e^{-V}\neq 0, 
\]
therefore $g^{\mu \nu }\nabla _\mu u\nabla _\nu u$ approaches infinity
rather than zero as $u\rightarrow 0^{+}$, and consequently the phase
boundary $\partial M_1$ is nonnull in the sense above. To check that the
example is really a regular solution to Ashtekar's equations, we make a
simple coordinate transformation 
\[
u=t-x,\quad V=t+x 
\]
and obtain from Eq.(12) that 
\begin{equation}
ds^2=e^{-V}[(U-U^{\prime })dt^2+2Udtdx+(U+U^{\prime })dx^2]+dY^2+dZ^2.
\end{equation}
Eqs.(11) and (13) imply 
\[
U-U^{\prime }\leq 0,\quad U+U^{\prime }\geq 0, 
\]
hence Eq.(14) is a standard formulation of the spacetime metric that can be
regarded as a regular solution to Ashtekar's constraint and evolution
equations 
\[
{\cal D}_aE_i^a=0,\ E_i^aF_{ab}^i=0,\ E_i^aE_j^bF_{abk}\epsilon ^{ijk}=0, 
\]
\[
\dot{E}_i^b=-i{\cal D}_a(\underline{N}E_j^aE_k^b)\epsilon ^{ijk}+2{\cal D}%
_a(N^{[a}E_i^{b]}), 
\]
\[
\dot{A}_b^i=i\underline{N}E_j^aF_{abk}\epsilon ^{ijk}+N^aF_{ab}^i, 
\]
where 
\begin{eqnarray}
E_1^a &=&\left( 
\begin{array}{c}
0 \\ 
-\frac 12[e^{-V}(U+U^{\prime })+1] \\ 
\frac i2[e^{-V}(U+U^{\prime })-1]
\end{array}
\right) ,  \nonumber \\
E_2^a &=&\left( 
\begin{array}{c}
0 \\ 
-\frac i2[e^{-V}(U+U^{\prime })-1] \\ 
-\frac 12[e^{-V}(U+U^{\prime })+1]
\end{array}
\right) ,  \nonumber \\
E_3^a &=&\left( 
\begin{array}{c}
1 \\ 
0 \\ 
0
\end{array}
\right) ,  \nonumber
\end{eqnarray}
and 
\[
A_a^i=0, 
\]
with lapse density and shift vector 
\[
\underline{N}=\frac{U^{\prime }}{U+U^{\prime }},\ \ N^1=\frac U{U+U^{\prime }%
},\ N^2=N^3=0. 
\]
Note that 
\[
\lim_{u\rightarrow 0^{+}}\underline{N}=1,\quad \lim_{u\rightarrow
0^{+}}N^1=0 
\]
as the degenerate region is approached.

It should be noted that the third derivative of the function $U(u)$ is not
continuous at the phase boundary $u=0$. However, the power 3 of $u$ in
Eq.(11) can be replaced by any real number greater than 3 to obtain the
desired differentiability.

\section{Discussions}

The idea that the phase boundary, $\partial M_1$, is always null is so
attractive that it seems intriguing to attribute the existence of the
counterexample presented in the previous section simply to the inappropriate
definition of the nullness of the phase boundary. As a matter of fact, the
definition used above has a fatal drawback: whether the phase boundary is
null depends upon the choice of the function which vanishes on the boundary.
Let $f$ and $\bar{f}$ be two distinct functions with $f=\bar{f}=0$ on the
phase boundary and $\nabla _\mu f|_{f=0}\neq 0$ and $\nabla _\mu \bar{f}|_{%
\bar{f}=0}\neq 0$, then there exists a function $\lambda $ on the boundary
such that $\nabla _\mu f|_{f=0}=\lambda \nabla _\mu \bar{f}|_{\bar{f}=0}$.
If $g^{\mu \nu }|_{f=0}$ were finite, then $g^{\mu \nu }\nabla _\mu f\nabla
_\nu f|_{f=0}$ would be equal to $\lambda ^2g^{\mu \nu }\nabla _\mu \bar{f}%
\nabla _\nu \bar{f}|_{\bar{f}=0}$, and hence $g^{\mu \nu }\nabla _\mu
f\nabla _\nu f|_{f=0}=0$ if and only if $g^{\mu \nu }\nabla _\mu \bar{f}%
\nabla _\nu \bar{f}|_{\bar{f}=0}=0$. However, since $g^{\mu \nu }|_{f=0}$ is
infinite, to judge the nullness of the phase boundary one has to calculate $%
g^{\mu \nu }\nabla _\mu f\nabla _\nu f$ and $g^{\mu \nu }\nabla _\mu \bar{f}%
\nabla _\nu \bar{f}$ in the nondegenerate side and then take the limit.
Since $g^{\mu \nu }\nabla _\mu f\nabla _\nu f\neq g^{\mu \nu }\nabla _\mu 
\bar{f}\nabla _\nu \bar{f}$ in general, there is no guarantee for the
equivalence of $\lim_{f\rightarrow 0}g^{\mu \nu }\nabla _\mu f\nabla _\nu
f=0 $ and $\lim_{\bar{f}\rightarrow 0}g^{\mu \nu }\nabla _\mu \bar{f}\nabla
_\nu \bar{f}=0$. If one chooses $f$ so that all $f=const.$ hypersurfaces in
the nondegenerate side are null, then $g^{\mu \nu }\nabla _\mu f\nabla _\nu
f=0$ everywhere in the nondegenerate side, hence the limit vanishes,
implying the nullness of the phase boundary. For instance, we could choose $%
f=ue^{(u-V)/3} $ in our example of Sec.3 to obtain this result. However, the
function $u$ in Sec.3 (playing the same role as $f$ here) was chosen
intentionally so that the hypersurfaces $u=const.$ are nonnull except for $%
u=0$, leading to the conclusion that $\lim_{u\rightarrow 0^{+}}g^{\mu \nu
}\nabla _\mu u\nabla _\nu u\neq 0$ and hence the same boundary becomes
nonnull. To save the attractive idea that the phase boundary, $\partial M_1$%
, is always null just as its image, $\phi [\partial M_1]$, it is necessary
to look for a suitable definition of the nullness of the boundary. (This
open question has been solved by the time when this paper is accepted [9].)

\acknowledgments

The authors would like to thank Prof. Ted Jacobson for his comments on the
original manuscript and his valuable suggestions. This work has been
supported by the National Science Foundation of China.

\end{document}